\documentclass[conference]{IEEEtran}
\IEEEoverridecommandlockouts
\usepackage{cite}
\usepackage{amsmath,amssymb,amsfonts}
\usepackage{algorithmic}
\usepackage{graphicx}
\usepackage{textcomp}
\usepackage{xcolor}
\usepackage{hyperref}
\usepackage{multirow}
\usepackage{makecell}
\usepackage{arydshln}
\usepackage{booktabs}
\def\BibTeX{{\rm B\kern-.05em{\sc i\kern-.025em b}\kern-.08em
    T\kern-.1667em\lower.7ex\hbox{E}\kern-.125emX}}
\begin{document}

\title{An Agent Framework for Real-Time Financial Information Searching with Large Language Models}

\author{
\IEEEauthorblockN{Jinzheng Li\textsuperscript{1}, Jingshu Zhang\textsuperscript{2}, Hongguang Li\textsuperscript{2}, Yiqing Shen\textsuperscript{3,*}}
\IEEEauthorblockA{
\textsuperscript{1}\textit{Fudan University}
\textsuperscript{2}\textit{JF SmartInvest Holdings}
\textsuperscript{3}\textit{Johns Hopkins University}\\
{\footnotesize \textsuperscript{*}Denotes corresponding authors (Email: yiqingshen1@gmail.com)}
}
}

\maketitle

\begin{abstract}
Financial decision-making requires processing vast amounts of real-time information while understanding their complex temporal relationships.
While traditional search engines excel at providing real-time information access, they often struggle to comprehend sophisticated user intentions and contextual nuances.
Conversely, Large Language Models (LLMs) demonstrate reasoning and interaction capabilities but may generate unreliable outputs without access to current data.
While recent attempts have been made to combine LLMs with search capabilities, they suffer from 
(1) restricted access to specialized financial data, 
(2) static query structures that cannot adapt to dynamic market conditions, 
and (3) insufficient temporal awareness in result generation.
To address these challenges, we present FinSearch, a novel agent-based search framework specifically designed for financial applications that interface with diverse financial data sources including market, stock, and news data.
Innovatively, FinSearch comprises four components: 
(1) an LLM-based multi-step search pre-planner that decomposes user queries into structured sub-queries mapped to specific data sources through a graph representation;
(2) a search executor with an LLM-based adaptive query rewriter that executes the searching of each sub-query while dynamically refining the sub-queries in its subsequent node based on intermediate search results;
(3) a temporal weighting mechanism that prioritizes information relevance based on the deduced time context from the user's query; 
(4) an LLM-based response generator that synthesizes results into coherent, contextually appropriate outputs.
To evaluate FinSearch, we construct \textit{FinSearchBench-24}, a benchmark of 1,500 four-choice questions across the stock market, rate changes, monetary policy, and industry developments spanning from June to October 2024.
Experiments demonstrate that FinSearch can outperform Perplexity Pro by 15.93\% with GPT4o, 14.06\% with LLama3.1-405B, and 21.6\% with Claude3.5-Sonnet.
The code is available at \href{https://github.com/eeeshushusang/FinSearch}{https://github.com/eeeshushusang/FinSearch}.
\end{abstract}


\section{Introduction}
\label{sec:intro}
Accessing accurate and timely information is important for making informed decisions in finance \cite{10.1145/3644713.3644743,kukreti2023stock}.
The dynamic nature of financial data, with stock prices updating by the minute and complex interdependencies between market factors, requires search methods that can efficiently integrate and analyze real-time information from diverse sources \cite{2311.15218}.
Traditional search engines, while providing access to real-time information, often fail to capture the nuanced relationships and temporal dependencies inherent in financial analysis \cite{10.1007/978-3-030-31624-2_4,anuyah2020empirical}.
Moreover, users struggle to express complex analytical needs through simple keywords, necessitating multiple iterations of query refinement, which can introduce delays in information access and potentially missing critical market signals. 
Furthermore, the need to initiate new search sessions for exploring related concepts creates a fragmented analytical process, hampering the comprehensive understanding of market dynamics and increasing the risk of overlooking important financial relationships \cite{10.5555/1630659.1630937}.

Large Language Models (LLMs) have enabled natural language understanding and generation, offering capabilities that complement traditional search functionalities \cite{2308.07107}.
Their ability to comprehend implicit context, extract key elements from natural language descriptions, and engage in multi-turn dialogues addresses many limitations of keyword-based search.
Additionally, LLMs excel at understanding user intent, automatically expanding queries, and refining responses based on user feedback. 
However, LLMs operate on static training data and therefore lack access to real-time information, leading to outdated analysis and potential hallucinations when addressing queries about current market conditions in finance \cite{2407.00128}.

Therefore, recent advances have highlighted the potential value of integrating search capabilities with LLMs to create AI agents capable of processing and responding to queries using real-time information \cite{rome2024ask}. 
MindSearch \cite{2407.20183} represents an initial step in this direction, introducing an AI agent framework for general search that systematically decomposes complex queries into detailed sub-queries.
However, its application to financial queries reveals three limitations. 
First, while MindSearch effectively utilizes general web search APIs such as \textit{Google} and Microsoft \textit{Bing}, it lacks integration with specialized financial data sources such as market data feeds, stock APIs, and financial news databases, which are essential for comprehensive financial analysis. 
Second, its query structure remains static throughout the search process.
In other words, once decomposed, sub-queries do not adapt based on intermediate search results, limiting their ability to respond to the dynamic nature of financial markets where new information can alter the relevance of previous queries.
Third, although MindSearch can process temporal information, it does not incorporate sophisticated temporal weighting mechanisms needed for financial applications where the importance of information varies based on its recency and relationship to market events.
While subsequent work MMsearch \cite{2409.12959} has extended MindSearch's capabilities to handle multi-modal scenarios, these fundamental limitations in handling financial data remain unaddressed.
Even commercial AI search products like \textit{Perplexity.ai}\footnote{\url{https://www.perplexity.ai}} encounter similar constraints when handling complex financial queries.

To address these limitations, we propose FinSearch, a novel search agent framework specifically designed for financial applications. 
At its core, FinSearch employs an LLM-based multi-step search pre-planner that decomposes user queries into structured sub-queries through a graph representation, with each sub-query mapped to specific financial API sources.
It enables coverage of diverse financial information streams while maintaining precise control over the search process.
Secondly, it distinguishes itself through previous work such as MindSearch through a search executor incorporating an LLM-based adaptive query rewriter that dynamically refines sub-queries based on intermediate search results, ensuring the search process remains responsive to emerging financial information. 
Moreover, FinSearch implements a novel temporal weighting mechanism that prioritizes information relevance based on the deduced time context from user queries, addressing the need for temporal awareness in financial analysis.
As shown in Table \ref{tab:Comparison}, FinSearch provides several unique capabilities compared to existing approaches.

\begin{table}[t!]
\centering
\caption{Comparison of AI search methods, highlighting the novel features of our proposed FinSearch framework against existing approaches. 
Code availability indicates if the implementation is publicly accessible. 
Cross-modal integration shows support for multiple data types including text and images. 
Dynamic query adaptation reflects the methods' ability to refine queries based on intermediate results. 
Time sensitivity indicates the capability to prioritize and weigh information based on temporal relevance.
Finance data availability indicates specialized access to financial data sources and APIs.
}\label{tab:Comparison}
\renewcommand{\arraystretch}{1.2} 
\resizebox{\columnwidth}{!}
{
\begin{tabular}{l|ccccc} 
\hline
\textbf{Methods} & 
\makecell[c]{\textbf{Code} \\ \textbf{Availability}} & 
\makecell[c]{\textbf{Cross-modal} \\ \textbf{Integration}} &
\makecell[c]{\textbf{Dynamic Query} \\ \textbf{Adaptation}} &  
\makecell[c]{\textbf{Time} \\ \textbf{Sensitivity}}&
\makecell[c]{\textbf{Finance Data} \\ \textbf{Availability}} \\ 
\hline
SearchAgent \cite{2407.20183}           & \textcolor{green}{$\checkmark$ }              & \textcolor{red}{$\times$ } & \textcolor{red}{$\times$ }    & \textcolor{red}{$\times$ }    & \textcolor{red}{$\times$ }                 \\
MindSearch \cite{2407.20183}    &\textcolor{green}{ $\checkmark$}     & \textcolor{red}{$\times$ }            & \textcolor{red}{$\times$ }       & \textcolor{red}{$\times$ }          & \textcolor{red}{$\times$ } 
\\ 
MMSearch \cite{2409.12959}                  &\textcolor{green}{ $\checkmark$}            &\textcolor{green}{$\checkmark$ }  & \textcolor{red}{$\times$ }          & \textcolor{red}{$\times$ }    & \textcolor{red}{$\times$ }     \\ 
\hdashline
Perplexity Pro  & \textcolor{red}{$\times$ }                    & \textcolor{green}{$\checkmark$ }   & \textcolor{red}{$\times$ }         & \textcolor{red}{$\times$ }       & \textcolor{red}{$\times$ }    \\
\hline
Ours                &\textcolor{green}{ $\checkmark$}     &\textcolor{green}{ $\checkmark$}         &\textcolor{green}{ $\checkmark$}     &\textcolor{green}{ $\checkmark$}  &\textcolor{green}{ $\checkmark$}  \\ 
\hline
\end{tabular}
    } 
\end{table}

The major contribution of this paper is four-fold. 
First, we introduce FinSearch, the first specialized search agent framework designed specifically for financial applications. 
It employs an LLM-based multi-step search pre-planner that decomposes complex financial queries into structured sub-queries with corresponding financial data sources.
Second, we propose a dynamic query adaptation mechanism.
Specifically, the search executor incorporates an LLM-based adaptive query rewriter that continuously refines sub-queries based on intermediate search results, ensuring the search process remains responsive to emerging market information. 
Third, we introduce a temporal weighting mechanism that prioritizes information relevance based on the deduced time context from user queries. 
Fourth, we develop \textit{FinSearchBench-24}, a benchmark comprising 1,500 multiple-choice questions that span diverse financial topics. 

\section{Methods}

\begin{figure*}[t!]
\centering
    \includegraphics[width=1\linewidth]{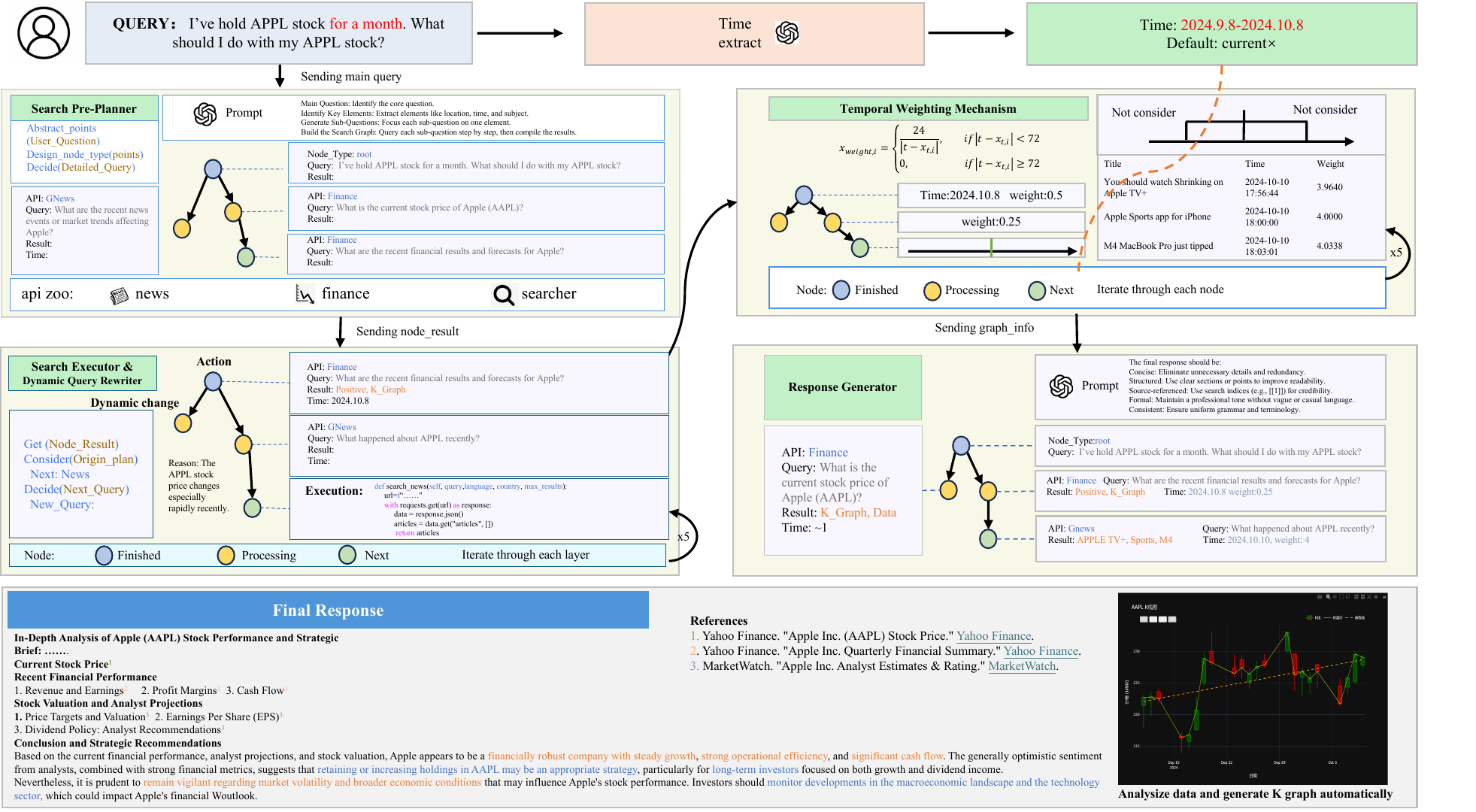}
        \caption{
Overview of the FinSearch. 
The user's financial query is first interpreted by a pre-planner that constructs a directed graph where nodes represent atomic search tasks connected to specific data sources. 
The search executor then traverses this graph, with a dynamic rewriter refining queries at each step based on emerging insights. 
The temporal mechanism evaluates the time-sensitivity of gathered information before the response generator creates a unified analysis incorporating both textual and visual elements.
}
    \label{fig:framework}
\end{figure*}

\subsection{Overview}
FinSearch is a search agent framework specifically designed for finance, that generates both textual analysis and visual outputs, such as k-line charts, in response to complex search queries.
FinSearch centers on a directed acyclic graph (DAG) structure, which we term a search graph, constructed by an LLM-based search pre-planner to represent the multi-step search process and store intermediate results.
Within the search graph, each node encapsulates a sub-query, which is a focused, atomic search component targeting specific aspects of the original complex query. 
These sub-queries are precisely mapped to designated financial data sources through selected APIs.
The directed edges between nodes represent both logical dependencies and temporal relationships between search components, creating a structured analytical pathway.
Following the initial graph construction, the search executor processes the sub-queries sequentially across nodes while dynamically optimizing subsequent queries for child nodes based on accumulated results to ensure that each successive search step builds upon and refines the insights gained from previous queries.
After completing the graph traversal, the temporal weighting mechanism applies time-aware weights to each node, ensuring that information is prioritized according to its chronological relevance to the user query. 
This temporal context is important for financial analysis, where the value of information often varies significantly with time.
The final component, the response generator, synthesizes the temporally weighted information collected across the search graph into comprehensive analytical responses with LLM. 
The overall framework of FinSearch is depicted in Fig.~\ref{fig:framework}.

\subsection{Search Pre-Planner}
The search pre-planner decomposes complex financial search queries into structured, executable sub-queries.
The process begins with semantic parsing, where an LLM analyzes the user query to extract key elements including temporal indicators $t$, corporate entities, and financial events. 
The pre-planner gives particular attention to temporal expressions, automatically converting relative time references (\textit{e}.\textit{g}., ``\textit{yesterday}'', ``\textit{last Friday}'') into precise dates to ensure temporal consistency throughout the subsequent search process.
Following semantic extraction, the pre-planner employs a human-inspired reasoning process to break down complex financial queries into sub-queries.
For instance, when analyzing a company's performance, the pre-planner might generate sequential sub-queries examining historical stock prices, recent earnings reports, market sentiment from news sources, and broader industry trends.
Correspondingly, each generated sub-query is precisely mapped to one of three specialized APIs: a news API for real-time information access, a search API for general market context, and a finance API for specialized financial data retrieval.

To orchestrate this multi-step search process, the pre-planner constructs a search graph, represented as a DAG $G(V_i, E_{ij})$.
The root node of $G$ contains the original query, while subsequent nodes represent increasingly specialized sub-queries arranged in a logical sequence that optimizes information gathering and analysis.
Each node $V_i$ in the search graph is characterized by a feature $x_i$ comprising five components, namely $x_{\text{query},i}$ representing the sub-query, $x_{\text{API},i}$ denotes the source API, $x_{\text{weight},i}$ serving as a temporal weight placeholder, $x_{\text{t},i}$ encoding the temporal context in the response, and $x_{\text{response},i}$ reserved for the response.
The directed edges $E_{ij}$ define execution dependencies between nodes according to $E_{ij}: V_i \rightarrow V_j$.

\subsection{Search Executor with Dynamic Query Rewriter}
The search executor processes the search graph generated from the pre-planner while continuously optimizing the search trajectory via the dynamic query rewriter.
For search execution, the executor traverses each node in the search graph sequentially. 
For each node, the executor generates API call codes based on the node's sub-query $x_{\text{query},i}$ and designated API $x_{\text{API},i}$.
The API execution process can be formalized as:
\begin{equation}
{x_{\text{response},i}, x_{\text{t},i}} = \texttt{execute}(x_{\text{query},i}, x_{\text{API},i})
\end{equation}
where \texttt{execute} represents the API calling function that returns both the search results $x_{\text{response},i}$ and their associated temporal information $x_{\text{t},i}$ directly from the API response.

The dynamic query rewriter enables adaptive search behavior that mirrors human analytical processes.
After each node execution, an LLM-based rewriter evaluates both the retrieved results and the current state of the planning graph to optimize subsequent sub-queries. 
This adjustment process can be formally represented as:
\begin{equation}
P(x_{\text{query},i}' | x_{\text{query},i}) = \texttt{rewriter}(x_{\text{query},i}, x_{\text{response},i}, G)
\end{equation}
where $x_{\text{query},i}$ denotes the current sub-query, $x_{\text{query},i}'$ for the revised sub-query, $x_{\text{response},i}$ represents the obtained result, and \texttt{rewriter} is the rewriter LLM that optimizes queries for subsequent iterations. 
This Markov chain-inspired approach allows FinSearch to adapt its search strategy based on accumulated insights continuously.

\begin{figure*}[t!]
\centering
\label{Fig.case} 
    \centering
    \includegraphics[width=0.97\linewidth]{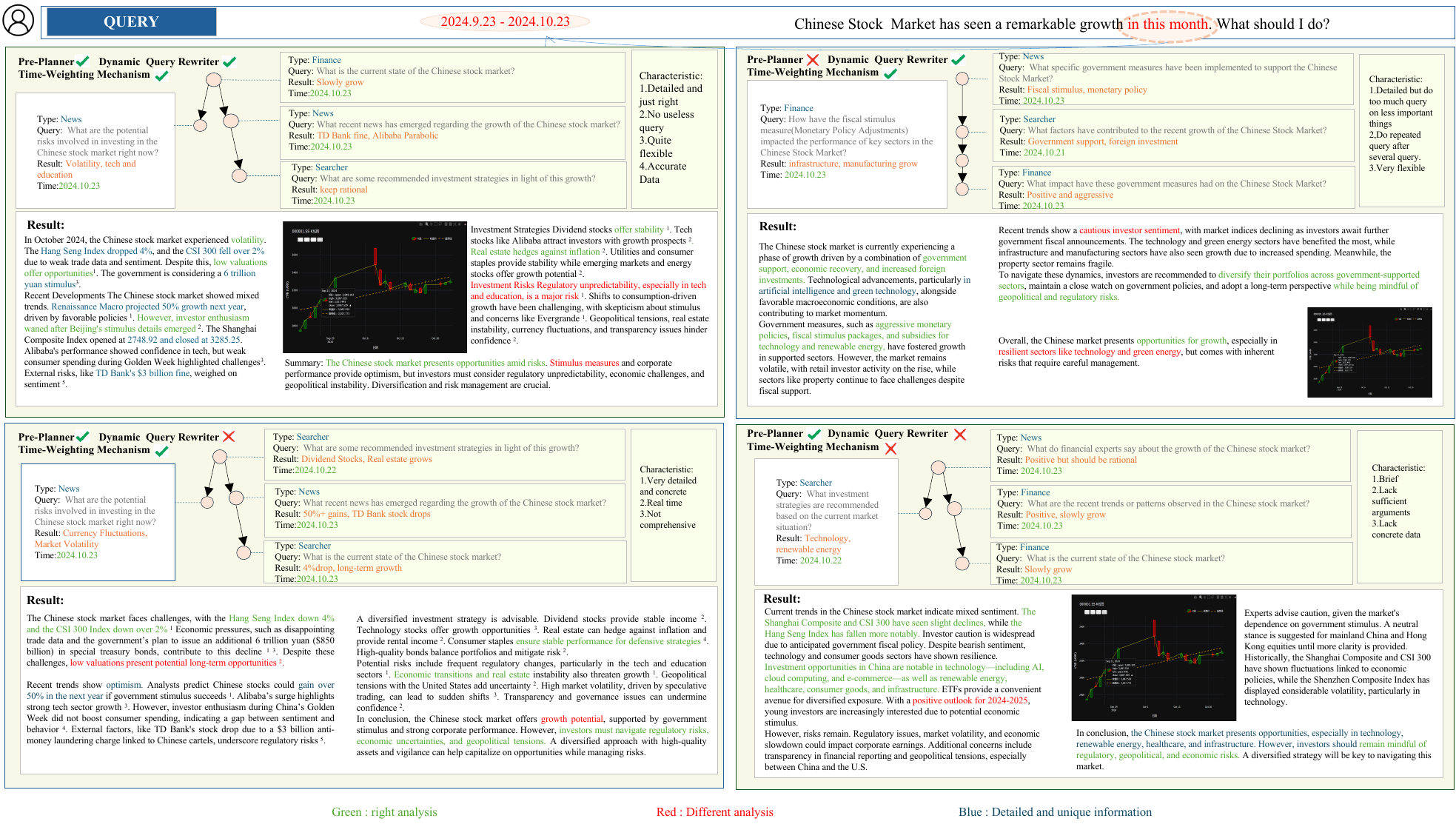}
\caption{
Case study for the ablation study.
We visualize FinSearch's analytical process across different configurations when analyzing Chinese stock market trends. 
It compares four distinct scenarios denoted by different colors: optimal configuration (green) showing accurate analysis aligned with market data, partial configuration without dynamic query rewriter (red) demonstrating divergent conclusions, and complete configuration with detailed temporal context (blue) providing comprehensive market insights.
Each panel illustrates FinSearch's query decomposition, information processing, and response generation, highlighting how different components contribute to analysis quality. 
The comparison reveals that the full configurations with all components enabled produce more nuanced, temporally-aware, and well-supported financial search results compared to configurations missing any features. 
}
\label{fig:ablation}
\end{figure*}

\subsection{Temporal Weighting Mechanism}
The temporal weighting mechanism ensures that recent information receives appropriate emphasis while historical context remains accessible when relevant \cite{app13085175}.
The mechanism operates by assigning temporal weights $x_{\text{weight},i}$ to each node based on its proximity to the query timestamp. 
For each node $V_i$ in the search graph, the temporal weight $x_{\text{weight},i}$ is computed using a time-decay function that considers the temporal distance between the information timestamp $t$ and the query timestamp $x_{\text{t},i}$:
\begin{equation}
x_{\text{weight},i} = 
\begin{cases}
\frac{24}{|t - x_{\text{t},i}|}, & \text{if } |t - x_{\text{t},i}| < 72 \\
0, & \text{if } |t - x_{\text{t},i}| \geq 72\\
\end{cases}.
\end{equation}
This weighting function implements a 72-hour temporal window, reflecting the rapid pace of financial markets where information beyond three days may become less relevant for many analytical purposes.
Within this window, the weight decreases linearly with temporal distance, ensuring that more recent information receives higher priority in the analysis.
The temporal weights are dynamically computed and updated during graph traversal, with each node's weight being stored in its feature vector component $x_{\text{weight},i}$.

\subsection{Response Generator}
The response generator synthesizes information from the temporally weighted search graph to produce comprehensive analytical outputs that combine textual analysis with visual representations by LLM. 
The generation process begins with information aggregation across all graph nodes, with each node's content being weighted according to its temporal relevance as determined by the temporal weighting mechanism. 
The generator performs reduplication to eliminate redundant information while preserving the unique temporal and analytical context of each data point. 
The textual component of the response adheres to strict citation protocols, ensuring that all insights are properly attributed to their respective sources. 
The generator maintains precise temporal context throughout the narrative, organizing information chronologically while emphasizing recent, highly weighted developments that are most relevant to the user's query.
For quantitative financial data, the generator incorporates interactive visualizations, particularly k-line charts obtained through the finance API, which display comprehensive trading information including daily opening, closing, high, and low prices represented through candlestick patterns. 
The final output integrates these textual and visual elements into a structured analytical report, with visualizations positioned strategically to support the narrative flow. 
Each component of the response is temporally contextualized, ensuring that users understand both the chronological sequence of events and their relative importance to the current analysis.

\begin{table*}[t!]
\centering
\renewcommand{\arraystretch}{1.5}
\caption{Performance comparison of search agent methods on \textit{FinSearchBench-24}, evaluating both accuracy and computational efficiency. 
The accuracy metrics show the percentage of correct answers ($\pm$ standard deviation). 
Time metrics represent the average seconds required per answer. 
Baseline refers to LLMs without search capabilities. 
SearchAgent and MindSearch represent general-purpose search frameworks, while Perplexity Pro is a commercial AI search solution. 
Our proposed FinSearch framework demonstrates superior accuracy across all tested LLMs while maintaining reasonable computational efficiency. 
Results marked with ``--'' indicate unavailable configurations. 
}
\setlength{\tabcolsep}{2.5pt} 
\label{tab:result}
\resizebox{\textwidth}{!}{ 
\begin{tabular}{c|c c c c c| c c c c c}
\toprule
\multirow{2}{*}{Method} & \multicolumn{5}{c|}{Accuracy (\%)} & \multicolumn{5}{c}{Time (s/answer)} \\ 
    \cline{2-11}
    \multicolumn{1}{c|}{} & 
    \multicolumn{1}{c}{\textbf{GPT-4o}} & 
    \multicolumn{1}{c}{\textbf{Llama3.1-405B}} & 
    \multicolumn{1}{c}{\textbf{Claude3.5-Sonnet}} &
    \multicolumn{1}{c}{\textbf{Deepseek}} &
    \multicolumn{1}{c|}{\textbf{Gemini-1.5-Flash}}&
    \multicolumn{1}{c}{\textbf{GPT-4o}} &
    \multicolumn{1}{c}{\textbf{Llama3.1-405B}} & 
    \multicolumn{1}{c}{\textbf{Claude3.5-Sonnet}}&
    \multicolumn{1}{c}{\textbf{Deepseek}}&
    \multicolumn{1}{c}{\textbf{Gemini-1.5-Flash}}\\
     \hline
    Baseline   & $36.13 \pm \scriptstyle{1.22} $& $38.13 \pm \scriptstyle{1.27}$ & $38.60 \pm \scriptstyle{1.28}$ & $34.07 \pm \scriptstyle{1.21}$ & $35.47 \pm \scriptstyle{1.22}$ & $3.87 \pm \scriptstyle{0.15}$ & $4.96\pm \scriptstyle{0.25}$ & $6.66 \pm \scriptstyle{0.21}$ & $14.50\pm \scriptstyle{0.27}$ & $5.04 \pm \scriptstyle{0.16}$
    \\ 
    \hline
    SearchAgent & $46.33 \pm \scriptstyle{1.30}$ & $43.80 \pm \scriptstyle{1.24}$ & $41.87 \pm \scriptstyle{1.29}$ & $44.27 \pm \scriptstyle{1.26}$ & $42.33 \pm \scriptstyle{1.28}$ & $\textbf{1.58}\pm\scriptstyle{0.06}$ & $\textbf{1.61}\pm\scriptstyle{0.07}$ & $\textbf{1.54}\pm \scriptstyle{0.06}$ & $\textbf{1.32}\pm\scriptstyle{0.04}$ & $\textbf{1.58}\pm \scriptstyle{0.04}$\\ 
    \hline
    MindSearch  & $52.40 \pm \scriptstyle{1.33}$ & $53.07 \pm \scriptstyle{1.34}$ & $53.60 \pm \scriptstyle{1.28}$ & $49.73 \pm \scriptstyle{1.32}$ & $51.53 \pm \scriptstyle{1.29}$ & $19.09 \pm \scriptstyle{0.39}$ & $14.82\pm \scriptstyle{0.45}$ & $17.93\pm \scriptstyle{0.39}$ & $27.01\pm \scriptstyle{0.58}$ & $20.14 \pm \scriptstyle{0.43}$\\ 
    \hline
    Perplexity Pro    & $60.27 \pm \scriptstyle{1.26}$ & $61.47 \pm \scriptstyle{1.28}$ & $56.67 \pm \scriptstyle{1.24}$ & -- & -- & $6.12 \pm \scriptstyle{0.26}$ & $3.94 \pm \scriptstyle{0.15}$ & $5.85 \pm \scriptstyle{0.22}$ & -- & -- \\ 
    \hline
    Ours       & $\textbf{76.20} \pm \scriptstyle{1.12}$ & $\textbf{75.53} \pm \scriptstyle{1.08}$ & $\textbf{78.27} \pm \scriptstyle{1.07}$ & $\textbf{72.33} \pm \scriptstyle{1.15}$ & $\textbf{74.87} \pm \scriptstyle{1.08}$ & $16.03 \pm \scriptstyle{0.43}$ & $14.55 \pm \scriptstyle{0.47}$ & $18.15 \pm \scriptstyle{0.35}$ & $29.31\pm \scriptstyle{0.70}$ & $17.74  \pm \scriptstyle{0.53}$ \\ 
    \bottomrule
    \end{tabular}
}
\end{table*}

\begin{table*}[t!]
  \centering
  \caption{
Ablation study evaluating the impact of FinSearch's functional components on performance across different LLM models using the \textit{FinSearchBench-24} dataset. 
Results demonstrate that both components contribute substantially to FinSearch's overall performance, with their combination yielding the highest accuracy across all tested models.
}
  \renewcommand{\arraystretch}{1.7}
  \label{tab:ablation}
  \resizebox{\textwidth}{!}{
  \begin{tabular}{c c| c c c c c| c c c c c}
   \toprule
    \multirow{2}{*}{\makecell[c]{Temporal Weighting\\Mechanism}} &\multirow{2}{*}{\makecell[c]{Dynamic Query\\Rewriter}} & \multicolumn{5}{c|}{Accuracy (\%)} & \multicolumn{5}{c}{Time (s/answer)} \\ 
    \cline{3-12}
    \multicolumn{2}{c|}{} & 
    \multicolumn{1}{c}{\textbf{GPT-4o}} & 
    \multicolumn{1}{c}{\textbf{Llama3.1-405B}} & 
    \multicolumn{1}{c}{\textbf{Claude3.5-Sonnet}} &
    \multicolumn{1}{c}{\textbf{Deepseek}} &
    \multicolumn{1}{c|}{\textbf{Gemini-1.5-Flash}}&
    \multicolumn{1}{c}{\textbf{GPT-4o}} &
    \multicolumn{1}{c}{\textbf{Llama3.1-405B}} & 
    \multicolumn{1}{c}{\textbf{Claude3.5-Sonnet}}&
    \multicolumn{1}{c}{\textbf{Deepseek}}&
    \multicolumn{1}{c}{\textbf{Gemini-1.5-Flash}}
    \\
    \hline
    $\times$ & $\times$       &$ 58.67 \pm \scriptstyle{\text{1.23}}$ &$ 56.40 \pm \scriptstyle{\text{1.33}}$ & $60.80 \pm \scriptstyle{\text{1.30}}$ & $55.27 \pm \scriptstyle{\text{1.26}}$ & $59.60 \pm \scriptstyle{\text{1.29}}$ & $20.17 \pm \scriptstyle{0.39}$ & $23.56 \pm \scriptstyle{0.31}$ & $26.64\pm \scriptstyle{0.27}$ & $39.19\pm \scriptstyle{0.36}$ & $\textbf{17.74} \pm \scriptstyle{0.29}$\\ 
    \hline
    \checkmark   & $\times$ & $72.87 \pm \scriptstyle{\text{1.14}}$ & $71.27 \pm \scriptstyle{\text{1.17}}$ & $73.67 \pm \scriptstyle{\text{1.13}}$ & $69.33 \pm \scriptstyle{\text{1.24}}$ & $72.40 \pm \scriptstyle{\text{1.16}}$ & $18.42 \pm \scriptstyle{0.37}$ & $24.15 \pm \scriptstyle{0.43}$ & $21.80 \pm \scriptstyle{0.48}$ & $35.24 \pm \scriptstyle{0.39}$ & $18.20 \pm \scriptstyle{0.23}$ \\ \hline
    $\times$ &\checkmark  & $62.33 \pm \scriptstyle{\text{1.22}}$ & $59.20 \pm \scriptstyle{\text{1.29}}$ & $65.40 \pm \scriptstyle{\text{1.22}}$ & $61.80 \pm \scriptstyle{\text{1.18}}$ & $63.47 \pm \scriptstyle{\text{1.21}}$ & $19.32\pm \scriptstyle{0.13}$ & $15.63 \pm \scriptstyle{0.27}$ & $19.51\pm \scriptstyle{0.39}$ & $\textbf{28.90}\pm \scriptstyle{0.19}$ & $16.25 \pm \scriptstyle{0.32}$ \\ \hline
    \checkmark &\checkmark  &$\textbf{76.20} \pm \scriptstyle{1.12}$ & $\textbf{75.53} \pm \scriptstyle{1.08}$ & $\textbf{78.27} \pm \scriptstyle{1.07}$ & $\textbf{72.33} \pm \scriptstyle{1.15}$ & $\textbf{74.87} \pm \scriptstyle{1.08}$ & $\textbf{16.03} \pm \scriptstyle{0.43}$ & $\textbf{14.55} \pm \scriptstyle{0.47}$ & $\textbf{18.15} \pm \scriptstyle{0.35}$ & $29.31\pm \scriptstyle{0.70}$ & $17.75  \pm \scriptstyle{0.53}$\\ 
    \bottomrule
    \end{tabular}  
}
\end{table*}

\section{Experiments}

\subsection{Benchmark Construction}
Following the previous design such as GTA \cite{2407.08713} and CFBenchMark \cite{2311.05812}, we developed \textit{FinSearchBench-24}, a benchmark comprising 1,500 multiple-choice questions spanning diverse financial topics spanning across June to October 2024.
Specifically, we implemented a four-phase construction process to ensure question quality, temporal relevance, and real-world applicability.
In the first phase, we collected current financial data from authoritative sources including government policy announcements, central bank communications, corporate earnings reports, and major financial news outlets, which ensured our benchmark reflected the latest market developments and regulatory changes across various financial sectors.
The second phase focused on structured question development. 
For each financial event or policy change, we created a series of questions tied to a specific timestamp. 
These questions were designed to assess three competencies: data retrieval accuracy (\textit{e}.\textit{g}., precise stock prices or economic indicators), information extraction (\textit{e}.\textit{g}., identifying key policy changes), and analytical understanding (\textit{e}.\textit{g}., evaluating market implications). 
In the third phase, we leveraged LLMs to scale our question-generation process. 
We developed detailed prompting templates that incorporated our question design principles and used them to generate additional questions based on our curated financial information. 
The final phase involved human validation by financial domain experts. 
Each question underwent review for technical accuracy, temporal consistency, and real-world relevance.

\subsection{Implementation Details}
We implemented FinSearch using Python 3.10.4 with multiple LLM backbones to evaluate performance across different model architectures. 
For our experiments, we tested five LLMs: GPT-4o, Llama3.1-405B, Claude3.5-Sonnet, Deepseek, and Gemini-1.5-Flash. 
Each model was accessed through its API using standard configurations without additional fine-tuning.
The search pre-planner and dynamic query rewriter components were implemented using identical prompting templates across all LLM backends to ensure a fair comparison.
For financial data retrieval, we integrated three categories of APIs: Yahoo Finance API for market data and stock information, NewsAPI for real-time financial news, and GoogleSearch API for general web search capabilities. 
The Yahoo Finance API was specifically configured to retrieve historical price data, company fundamentals, and market indicators with a sampling rate of one minute for real-time data.
The graph nodes were designed as Python objects.

\subsection{Results}
The experimental results in Table \ref{tab:result} demonstrate that our proposed FinSearch outperforms existing search agent methods across multiple LLM architectures. 
FinSearch achieves superior accuracy compared to baseline approaches and state-of-the-art search agents while maintaining reasonable computational efficiency.
When examining accuracy metrics, FinSearch consistently delivers the strongest performance across all tested LLM backends. 
With GPT-4o, FinSearch achieves 76.20\% accuracy, representing a 15.93\% improvement over Perplexity Pro (60.27\%) and a substantial 23.80\% gain over MindSearch (52.40\%). 
Similar patterns emerge with other LLM architectures, \textit{e}.\textit{g}. using Llama3.1-405B, FinSearch attains 75.53\% accuracy compared to Perplexity Pro's 61.47\%, while with Claude3.5-Sonnet, FinSearch reaches 78.27\% accuracy versus 56.67\% for Perplexity Pro. 
The framework also demonstrates robust performance with newer LLM architectures, achieving 72.33\% and 74.87\% accuracy with Deepseek and Gemini-1.5-Flash respectively, though comparative data for Perplexity Pro is not available for these models.
Regarding computational efficiency, FinSearch maintains reasonable processing times despite its more sophisticated search and analysis capabilities. 
While SearchAgent shows the fastest processing times (approximately 1.5 seconds per answer), it achieves this at the cost of lower accuracy. 
FinSearch's average processing time ranges from 14.55 to 29.31 seconds per answer across different LLMs, which is comparable to MindSearch (14.82 to 27.01 seconds) but longer than Perplexity Pro (3.94 to 6.12 seconds). 
This additional computational cost is justified by the substantial gains in accuracy, particularly for time-sensitive financial applications where precision is paramount.
The baseline LLM implementations without search capabilities perform notably worse across all metrics, with accuracy rates between 34.07\% and 38.60\%, underscoring the importance of integrating real-time search capabilities for financial analysis tasks.

\subsection{Ablation Study}
To evaluate the individual contributions of FinSearch's core components, we conducted an ablation study examining the impact of the temporal-weighting mechanism and dynamic query rewriter. 
Table \ref{tab:ablation} presents the results across different LLM architectures, demonstrating that both components contribute to the framework's overall performance.
The temporal-weighting mechanism emerges as the more impactful component, yielding substantial accuracy improvements when implemented alone. 
For instance, with GPT-4o, incorporating only the temporal-weighting mechanism increases accuracy from 58.67\% to 72.87\%, representing a 14.20\% improvement. 
Similar gains are observed across other LLM architectures, with improvements ranging from 12.80 to 14.87 \% points. 
This confirms our hypothesis that temporal context plays an important role in financial information processing.
While showing a more modest individual impact, the dynamic query rewriter still provides meaningful performance improvements. 
When implemented in isolation, it improves accuracy by 3.66 \% with GPT-4o (from 58.67\% to 62.33\%) and demonstrates consistent gains across all LLMs. 
This suggests adaptive query refinement helps capture more relevant information during the search process.
Most notably, combining both components produces synergistic effects, achieving the highest accuracy across all LLM configurations. 
With both components active, FinSearch reaches a peak performance of 78.27\% accuracy using Claude3.5-Sonnet, representing a 17.47\% improvement over the baseline configuration without either component. 
This synergy indicates that temporal awareness and dynamic query adaptation work complementarily to enhance search effectiveness.
We provide a case study in Fig. \ref{fig:ablation}.
Regarding computational efficiency, the full configuration with both components demonstrates improved processing times compared to partial implementations in most cases. For example, with Claude3.5-Sonnet, the complete system processes queries in 18.15 seconds on average, compared to 26.64 seconds for the baseline configuration. This suggests that the additional components help focus the search process more effectively, potentially reducing unnecessary computational overhead.

\section{Conclusion}
This paper introduces FinSearch, a novel search agent framework specifically designed to address the unique challenges of financial information retrieval and analysis. 
FinSearhc's innovations include an LLM-based multi-step search pre-planner that decomposes complex financial queries through graph representation, a dynamic query rewriter that adaptively refines searches based on intermediate results and a temporal weighting mechanism that ensures appropriate prioritization of time-sensitive financial information. 
Our extensive experimental evaluation on the newly developed \textit{FinSearchBench-24} dataset demonstrates FinSearch's superior performance across multiple LLM architectures. 
Looking forward, this work opens several promising directions for future research. 
FinSearch could be extended to support additional financial data sources and analysis types, while the temporal weighting mechanism could be refined to better handle different time horizons and market conditions. 
We believe FinSearch represents a step forward in developing specialized search agents for financial applications, demonstrating how careful integration of LLM capabilities with domain-specific requirements can yield substantial improvements in search accuracy and relevance.

\bibliographystyle{IEEEbib}
\bibliography{main.bib}

\end{document}